# Research Traceability using Provenance Services for Biomedical Analysis

Ashiq Anjum, Peter Bloodsworth, Andrew Branson, Irfan Habib, Richard McClatchey,
Tony Solomonides and the neuGRID Consortium
Centre for Complex Cooperative Systems, Univ. of the West of England, Bristol UK

**Abstract:** We outline the approach being developed in the neuGRID project to use provenance management techniques for the purposes of capturing and preserving the provenance data that emerges in the specification and execution of workflows in biomedical analyses. In the neuGRID project a provenance service has been designed and implemented that is intended to capture, store, retrieve and reconstruct the workflow information needed to facilitate users in conducting user analyses. We describe the architecture of the neuGRID provenance service and discuss how the CRISTAL system from CERN is being adapted to address the requirements of the project and then consider how a generalised approach for provenance management could emerge for more generic application to the (Health)Grid community.

## 1. Introduction

In the HealthGrid community great emphasis has been placed on the provision of suitable Grid infrastructures to support biomedical researchers for the purposes of, amongst others, data capture, image analysis, the processing of biomedical algorithms and the sharing of diagnoses [refs]. Many projects have reported on the customisation of Grid middleware such as gLite and GRIA [1] and on the provision of access through portals to data distributed on the Grid between centres of biomedical research, for example those studying degenerative brain diseases. Lately neuroscience projects such as NeuroGrid [2], NeuroLog [3] and neuGRID [4] have considered providing services based on service-oriented architectures to facilitate the complex studies required to analyse MRI (and CT) images including querying and browsing data samples and specifying and executing workflows (or pipelines) of algorithms required for neurological image analysis. To date few (if any) have considered how such analyses can be tracked over time, between researchers and over varying data samples and analysis pipelines.

One area of increasing computer science research in recent years has been that of so-called 'provenance' data capture and management. Here provenance means the history, ownership and usage of data in some domain of interest, for example the tracking of engineering samples in the construction of the Airbus or the logging of data and process execution in the study of High Energy Physics (HEP) experiments [5]. Many approaches have been followed for provenance management and recently the Open Provenance Model effort [10] has attempted standardisation of provenance models and their access. Researchers are now looking to ((re-)engineer their ad-hoc provenance management systems to follow OPM guidelines and encouragement is given for its application in other domains including the biomedical.

In this paper we outline the approach being developed in the neuGRID project [6] to use provenance management approaches (to be based on the OPM) for the purposes of capturing and preserving the provenance data that emerges in the specification and execution of (stages in) analysis pipelines and in the definition and refinement of data samples used in studies of Alzheimer's disease (AD) [7]. The neuGRID project is adapting a provenance tracking system called CRISTAL [8] which has been developed at CERN, Geneva to manage the construction of large-scale HEP detectors for the Large Hadron Collider, for the purposes of tracking neurological analyses of AD. In the next section we provide the background for this study, followed in section 3 by a discussion of the domain requirements for provenance capture using a tangible use-case from the neuGRID project. Later we consider the architecture of the neuGRID provenance service and discuss how CRISTAL is being adapted to address the requirements of the project and then consider how a generalised approach for provenance management could emerge for more generic application to the (Health)Grid community. The paper concludes with discussion of future possible research including how a provenance repository could act as a knowledge resource for providing guided assistance to clinical researchers in their biomedical analyses.

## 2. Data Provenance in Research Environments

In clinical research environments analyses or tasks can be expressed in the form of workflows (elsewhere called pipelines). A scientific workflow is a step-wise formal specification of a scientific process; one example of a workflow is one which represents, streamlines, and automates the steps from dataset selection and integration, computation and analysis, to final data product presentation and visualization. (An example of a neuGRID workflow is discussed in the next section). A workflow management system supports the specification, execution, re-run, and monitoring of scientific processes. Due to the complexity of scientific (or biomedical) workflows researchers require a means of tracking the execution of specified workflows to ensure that important analyses are accurately (and reproducibly) followed. Currently this is carried out manually and can be error-prone.

As this is currently a manual process, the complexity of biomedical analysis processes may inadvertently lead to an execution failure or undesired execution results. These may include, amongst others, incorrect workflow specifications, inappropriate links between pipeline components, execution failures because of the dynamic nature of the resources. A real problem in this scenario is tracking faults as and when they happen. This is mainly due to the absence of an information capturing mechanism during the workflow specification, distribution and execution. Hence a user is often unable to track errors in past neuroimaging analyses. This may in turn lead to a loss of user control or repetition of errors during subsequent analyses. Users may be prevented from being able to:

- Reconstruct a past workflow or parts of it to view the errors at the time of specification.
- Validate a workflow against a reference specification.
- Validate workflow execution results against a reference dataset.
- Query information of his interest from the past analysis.
- Compare different analyses.
- Search annotations associated with a pipeline or its components for future reference.

The benefit of managing specified workflows over time is that they can be refined and evolved by users and can ultimately reach a level of maturity and dependability. Users consequently need to gather information about (versions of) workflow specifications that may have been gathered from multiple users together with whatever results or outcomes were generated and then to use this so-called 'provenance data' as the drivers for improved decision making. The execution of such improved workflows may generate additional outcome provenance data which must again be captured, managed and made accessible to researchers to inform and facilitate future analyses and to provide traceability of their research data and algorithms. This provenance data may, over time, become an important source of acquired knowledge as the nature of the analyses evolve; the provenance data store will thus essentially become a knowledge base for researchers. Users can invoke services to automatically monitor and analyze large provenance databases to return statistical results that match criteria as set by the end user. This should provide efficient and dependable problem solving functionality in a controlled decision support system for the researchers.

The aim of the neuGRID project is to provide a user-friendly grid-based e-infrastructure plus a set of generalised infrastructure services that will enable the European neuroscience community to carry out research that is necessary for the study of degenerative brain diseases. In the neuGRID project a provenance service has been designed and implemented that is primarily intended to capture the workflow information needed to populate a project-wide provenance database. The provenance service will keep track of the origins of the data and its evolution between different stages of research analyses. The provenance service will allow users to query analysis information, to regenerate analysis workflows, to detect errors and unusual behaviours in past analyses and to validate analyses. The service will support and enable the continuous fine-tuning and refinement of the workflows in the neuGRID project by capturing:

- Workflow specifications.
- Data or inputs supplied to each workflow component.
- Annotations added to the workflow and individual workflow components.
- Links and dependencies between workflow components.
- Execution errors generated during analysis.
- Output produced by the workflow and each workflow component.

## 3. Requirements Analysis: Provenance in neuGrid

The user requirements gathering process in neuGRID involved working closely with the clinical researcher community. Meetings focused initially on the description of high-level stories and usage patterns that would later be used to cross-check system functionality during final system testing. As these were produced a range of use-cases were created and then prioritised. This provided a clear framework on which more detailed individual requirements could be based and this has been of benefit in terms of describing the project and ensuring that important components are not overlooked. This also led to a clear hierarchical conceptual framework being identified that linked high-level stories to more finely grained use-cases and to individual users requirements. The primary focus of this work was on the production of easily understandable models that are meaningful to both clinical researchers and software developers.

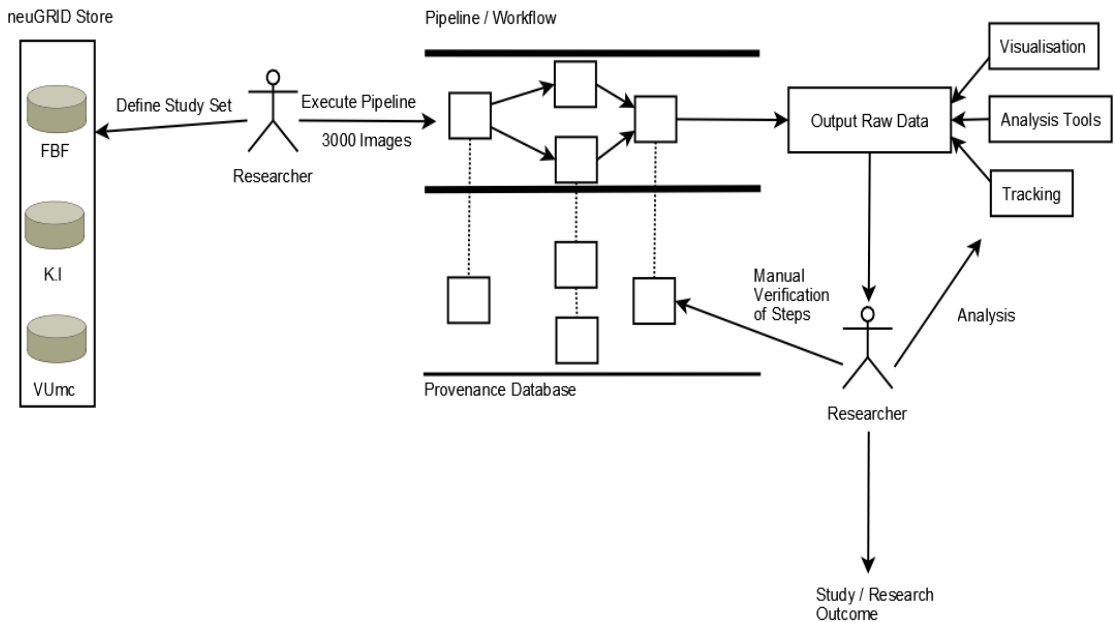

Figure 1: Validation of Results using Provenance Data Story

The following story (see figure 1) was identified during the requirements gathering process and illustrates the valuable role that accurate provenance information can play in the research process. Consider the situation in which a given workflow yields some surprising and possibly significant results. A researcher wishes to confirm that the results are accurate and to identify any mistake that nay have been made. By analysing all the intermediary image sets and workflow execution logs the user is able to manually verify that the results were incorrect. It is found that the error was due to a specific group of images interacting badly within the workflow. The user annotates the workflow so that other users are warned if they attempt a similar analysis. A data provenance system would enable the capture of the workflow specification, the outcomes of each run of that workflow on a specified data sample and the annotation provided by researchers on the execution of that workflow. Consequently the provenance database would begin to act as a shared knowledge-base for the community of researchers.

Neurological science relies heavily on the use of statistical analysis techniques in order to process the output from a workflow and thereby test a given research hypothesis. A key factor in being able to draw meaningful conclusions from data is the size of the study sample. Generally speaking, the greater the size of the study set the more confidence that can be given to the results that are produced. It may also be that a larger sample size will allow more precise questions to be asked which can lead to the discovery of new correlations between variables. A potential problem that arises when working with large numbers of scans is that the highly sensitive imaging processing algorithms may often fail in a proportion of cases. Such errors may have a significant impact on the analysis results and potentially allow incorrect conclusions to be reached. Such information may play an important role in the development of new treatments and the evaluation of the efficacy of experimental drugs and it is therefore important that errors are discovered before research outcomes are published. A provenance database would track the evolution of data samples as produced by the researchers and therefore robust data provenance is a high priority for researchers.

neuGRID is an example of an infrastructure that has been designed to provide researchers with a shared set of facilities through which they can carry out their research. At the heart of the platform is a distributed computation environment which is designed to efficiently handle the running of image processing workflows such as the cortical thickness measuring algorithm, CIVET [9]. This is not enough on its own however, as users require more than simply processing power. They need to be able to access a large distributed library of data and to search for a group of images with which they will work. A set of common image processing workflows is also necessary within the infrastructure for users to work with. A significant proportion of clinical research involves the development of new workflows and image analysis techniques. The ability to edit existing and to construct new workflows using established tools is therefore important to researchers. Another vital aspect is the traceability of the analysis data that is produced using a workflow. Researchers need to be able to examine each stage in the processing of an analysis workflow in order to confirm that it is accurate. Overall users expect a well tailored research infrastructure to support them in their research. Provenance information plays a crucial role in achieving this by bringing together and storing information regarding how individual users interact with the underlying data infrastructure.

**4. Provenance Service: Architecture and Components**

The design of the Provenance Service has been derived from the initial set of requirements that have been outlined in section 3. Once these requirements have been met, optional features were added to the Provenance Service as described in section 3. Figure 2 shows the functionality the Provenance Service will provide and where it is positioned in the neuGRID architecture in which a three layer approach has been followed. The client services will invoke the Provenance Service to initiate provenance capture. This also includes the workflow specification provenance. The actual provenance information will be retrieved from the underlying infrastructure where data that has been generated during an analysis resides. Later the analysis history of pipelines will be provided by the querying interfaces of the provenance service.

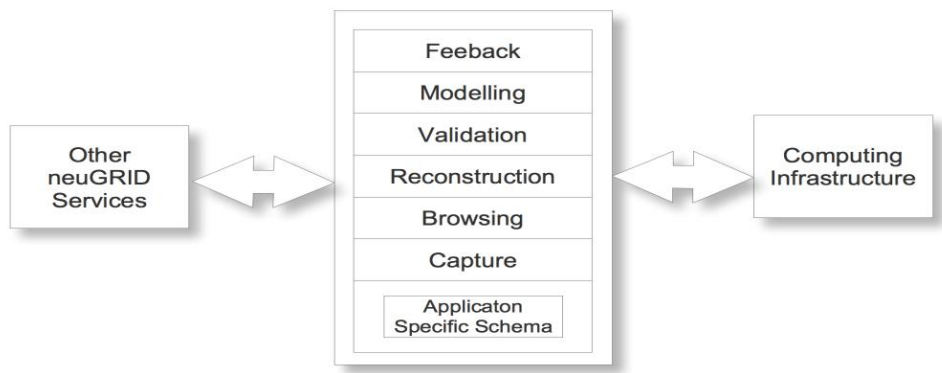

Figure 2: Provenance Service Design

The design and implementation of the provenance service will be OPM [10] compliant. Compliance with the OPM will enable interoperability between various OPM compliant provenance systems to share analysis history and experiences across projects and to enable reuse. One of the main issues with non-standardised provenance platforms is that the tools to query and analyse provenance are platform specific. Moreover, ad-hoc methods are employed to transfer provenance information between systems. Often this process results in the transfer of incomplete or inaccurate provenance information. The OPM does not define a provenance representation schema, however it defines schemas such as an XML schema used only for mapping concepts and transferring PM structures between systems. The OPM also enables developers to develop and share tools that work across systems.

To implement OPM compliance, there are essentially three steps:
- Describe an application specific schema. This process maps OPM concepts to their corresponding instantiations.

- Describe provenance in terms of OPM Graphs. An OPM Graph is a set of connected processes, artefacts, agents and other entities that demonstrates the provenance of a specific entity from its origin to its final state.
- Provide the ability to import and export OPM Graphs. Importing provenance or exporting provenance between OPM compliant systems requires the ability to import and export OPM Graphs described in the standardised OPM XML Schema [11].

Using the Provenance Service all provenance information that will be captured will be stored in an Open Provenance Model compliant schema. An application specific schema and database access mechanism needs to be defined for each client application. This schema and the database is the only component in the provenance service that changes with each new application to provide a fine grained querying and storage mechanism. The provenance information can be stored on remote databases, which will be accessed through SOAP or similar protocols; such support needs to be implemented as discussed in section 5.

Following database design, the next step is to capture the provenance at specification and executions stages. Provenance capture will be a key component of the provenance service, as its interface will allow a user/client service to store specification-level provenance and execution logs into the provenance database. The capturing component of the provenance service will be able to capture and store (in an application specific provenance data) the following parameters:

- Workflow descriptions and version information.
- The head node of the workflow.
- The input data supplied to the head node.
- A script or process associated with a workflow node.
- The type of a workflow node i.e. single process node or composite node.
- The successors of a workflow node.
- The predecessors of a workflow node and input data supplied to it.
- Metadata associated with each workflow node.

Once the information has been stored into a provenance database, it needs a mechanism to be queried and browsed by the application users. The browsing component will be built on the top of the provenance database, which will serve as a utility for the users to browse the past pipeline traces. Browsing is not itself a core component of the provenance service, rather it is an important project requirement which will help users to interact with and use the provenance database in a simplified way. A pipeline comprises a start node, tasks or actors, successors and predecessors of an actor, links/relations among actors, input data and files supplied to each actor, and a final output of complete pipeline. The provenance database will store each of the pipeline's constituent parts. This will enable a user to retrieve a complete pipeline from the provenance store and also examine sections of it with the appropriate dependencies. This is important, as it will allow users to examine the various stages in the pipeline creation process (perhaps for pipeline debugging) even if they cannot remember each and every step that they originally took. The reconstruction APIs will help a user to reconstruct a pipeline or part of it for different purposes such as:

- Observing the pipeline creation process in past
- Re-executing a pipeline or part of it
- Modifying a pipeline and storing it with a different version

Users will specify neuroimaging pipelines by combining different analysis algorithms. The pipelines will be 'gridified' and the algorithms will then be executed over the Grid. The pipeline designer/creator may accidentally define inappropriate links between different components. A change to an analysis algorithm, residing on grid sites, may not always propagate to the user end. Therefore at the time of creating pipeline its author would not be aware of any change in the logic of an algorithm. In these situations a user may receive corrupted or outdated execution results. The validation component of the provenance service will enable a user to verify a current pipeline specification against a reference blueprint. It will also allow a user to verify the results of an execution using a reference dataset. This type of validation will be performed in two ways: a) re-executing algorithms in the pipeline and then comparing the results of the execution with the reference/expected output will perform an online validation of the results or b) offline validation will verify the results of an already executed pipeline, in the provenance database, with a reference dataset.

The modelling component will analyse and group the provenance information that is stored in the provenance database. This will lead to the identification of groups of data that represent different patterns that occur during pipeline specification and execution behaviours. The modelling process could harness machine learning algorithms in order to categorize information segments that are present in the provenance database. Different techniques of evolutionary computing could be applied to identify new information chunks and thus modelling would become a dynamic and evolving process. The feedback component could use the modelling information to provide users with useful information that is driven by the modelling process. The role of this component would be similar to a decision support system, which would help in pipeline re-construction and the planning process. This may, for example, suggest the use of a specific analysis algorithm, process, an experience or an observation that has been saved as a best practice. Both the modelling and feedback processes would be stochastic and their efficiency would improve as the provenance database grows.

## 5. CRISTAL as a Provenance Management Platform

CRISTAL is a data and workflow tracking (i.e. provenance) system which is being used to track the construction of large-scale experiments such as CMS at the CERN Large Hadron Collider (LHC). It is a process modelling and provenance capture tool that addresses the harmonisation of processes by the use of the CRISTAL Kernel so that multiple potentially heterogeneous processes can be integrated with each other and have their workflows tracked in the database. Using the facilities for description and dynamic modification in CRISTAL in a generic and reusable manner, CRISTAL is able to provide modifiable and reconfigurable workflows. It uses the so-called description-driven nature of the CRISTAL models to act dynamically on process instances already running and can thus intervene in the actual process instances during execution. These processes can be dynamically (re-)configured based on the context of execution without compiling, stopping or starting the process and the user can make modifications directly and graphically of any process parameter, while preserving all historical versions so they can run alongside the new. In the provenance service, we have used CRISTAL to provide the provenance needed to support neuroscience analyses and to track individualised analysis definitions and usage patterns thereby creating a knowledge base for neuroscience researchers. This section describes how CRISTAL fits into the overall neuGRID architecture to capture and coordinate provenance data. The subsequent sections explain the implementation details of CRISTAL and highlight how provenance is captured, modelled, stored and tracked through the course of an analysis.

As shown in figure 3, the interaction starts with the authoring of a pipeline, which the user wants to execute on the Grid. Authoring can be carried out via several tools, the prototype being implemented in neuGRID, uses Kepler [12] and the LONI Pipeline [13] as examples of authoring environments. The workflow specification is enriched by including provenance actors for provenance collection. The Pipeline Service translates the workflow specification into a standard format and plans the workflow. The planned workflow, as shown in figure 3, is forwarded to the CRISTAL enabled provenance service which then creates an internal representation of this workflow and stores the workflow specification into its schema. This schema has sufficient information to track the workflow during subsequent phases of a workflow execution. The workflow activity is represented as a tree like structure (actually a directed acyclic graph to be accurate) and all associated dependencies, parameters, and environment details are represented in this tree. The schema also provides support to track the workflow evolution and the descriptions of derived workflows and its constituent parts are related to the original workflow activity.

The provenance service provides a provenance-aware workflow instantiation engine. The workflow is broken into its constituent jobs and CRISTAL takes care of the jobs, their dependencies and the order in which these should be executed to complete a workflow. CRISTAL coordinates the whole job execution process and the jobs wait inside the CRISTAL premises if their dependent tasks are in execution. The workflow is instantiated in a task-by-task manner by CRISTAL. These tasks are submitted to Grid for execution and the results and logs are retrieved to populate a provenance structure. CRISTAL is unaware of how the actual scheduling, task allocation and execution is carried out in the underlying Grid infrastructure. All of these operations are performed independently from CRISTAL. The information stored in the provenance structure can be interactively queried by users.

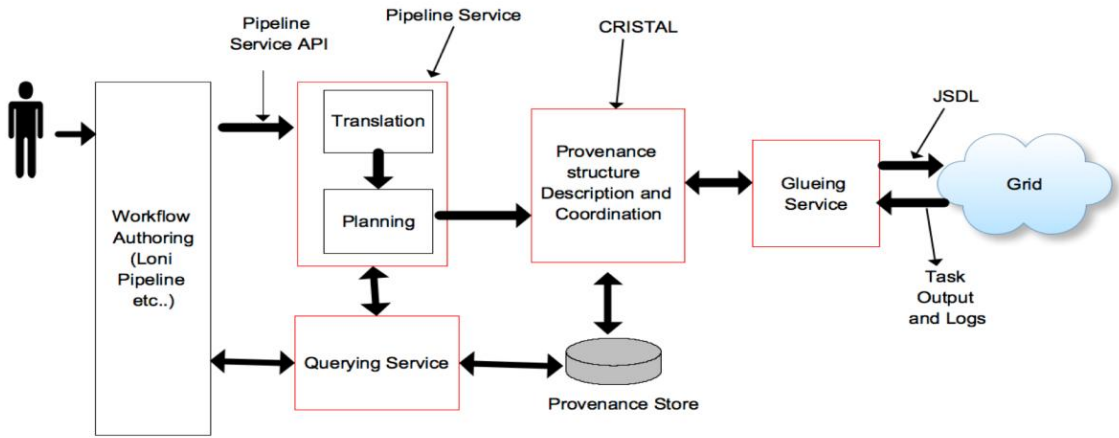

Figure 3: CRISTAL enabled Provenance Service Architecture

As shown in figure 4, the Pipeline Service will forward a concrete planned pipeline to the Provenance Service. At this stage a wrapper will be invoked that will initialise the appropriate structures required for provenance tracking. The wrapper will first create agent, outcome, activity and collection descriptions. Agents in the CRISTAL model execute specific activities. In neuGRID, and in computing environments in general, a compute element is an agent since it executes the workflow activities. Outcomes are the outcomes of individual activities. Activity descriptions contain descriptions of the atomic tasks in a workflow. All these descriptions constitute an "Item" in CRISTAL.

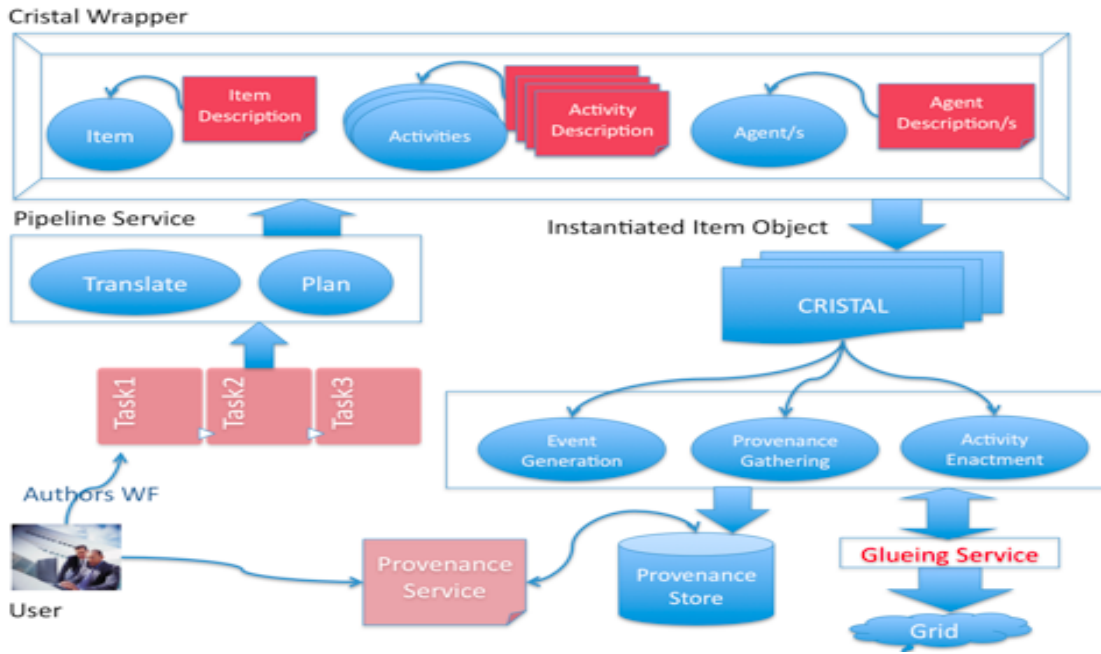

Figure 4: CRISTAL Architecture and Components

The Item represents the entire workflow and all the associated descriptions and activities. Every time a workflow is executed an event is generated which stores the outcome of the workflow. An item will have several events in case workflows are executed a number of times. To execute a workflow a new event is generated. The event generation starts the execution of the workflow by initialising an activity object. The activity object at first initialises a state machine of the workflow. The state machine tracks progress of an activity's state while it is being executed. Activities can be described with a certain number of states such as "Suspended", "Interrupted" etc. The state machine iteratively executes all activities in a workflow and stores the corresponding provenance.

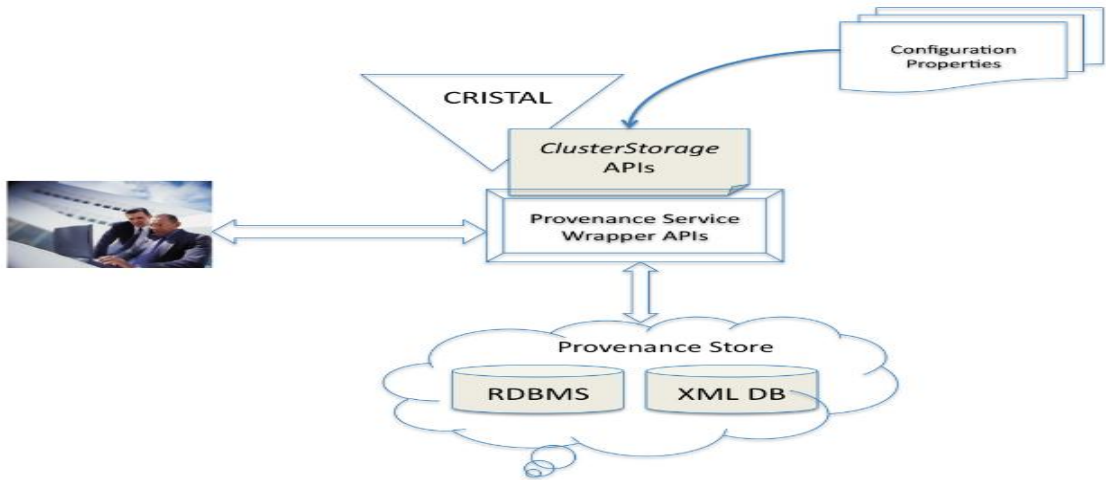

Figure 5: Provenance Service Structure

CRISTAL also has pluggable data storage to store and retrieve provenance information. This can be extended by replacing its default storage mechanism that stores XML files that can be queried via OpenLDAP. The key interface for this purpose is ClusterStorage. This interface uses a configuration file to connect to different relational databases. As shown in figure 5, ClusterStorage provides functions to access databases and these functions must be overridden for using a specific database. CRISTAL uses ClusterStorage to store properties, events, views, outcomes, workflows and collections for each Item. The Item itself contains "Paths" to these elements, which CRISTAL accesses through the ClusterStorage API. In the context of neuGRID project, ClusterStorage interface will be wrapped by Provenance Service APIs, which will provide extended functionality for recording and querying provenance information.

**6. The Way Forward**

In neuGRID each user will have separate authentication and authorisation credentials. Therefore the issues such as granularity of authorisation and synchronisation of the CRISTAL data security with the security deployed in the rest of neuGRID need to be further explored. The current provenance reconstruction mechanism in CRISTAL is not sufficient to enable the scientists to reconstruct their workflows. The reconstruction process should help in observing the pipeline creation process, re-executing a pipeline or part of it and modifying a pipeline and storing it with a different version. The current schema and database access mechanism needs to be refined to provide a fine grained querying and storage mechanism. The provenance information may be stored on remote databases, which will have to be accessed through SOAP or similar protocols and such support is necessary in CRISTAL.

In the long run we intend to research and develop an analysis module in CRISTAL. This will enable applications to learn from their past executions and improve and optimize new studies and processes based on the previous experiences and results. Using machine learning approaches, models can be formulated that can derive the best possible optimization strategies by learning from the past execution of experiments and processes. These models will evolve over time and will facilitate in the decision support in designing, building and running the future processes and workflows in a domain. A provenance analysis mechanism will be built on top of the OPM compliant data that has been captured in the CRISTAL system. It will employ approaches to learn from the data that has been produced, find common patterns and models, classify and reason from the information accumulated and present it to the system in an intuitive way. This information will be delivered to users while they work on new processes or workflows and will be an important source for their future decision making.

**7. Conclusions**

In this paper we outlined the approach being developed in the neuGRID project to use provenance management approaches for the purposes of capturing and preserving the provenance data that emerges in the specification and execution of (stages in) analysis pipelines and in the definition and refinement of data samples used in studies of Alzheimer's disease (AD). In the neuGRID project a provenance service has been designed and implemented that is primarily intended to capture the workflow information needed to populate a project-wide provenance database. The provenance service can keep track of the origins of the data and its evolution between different stages of research analyses. The provenance service can allow users to query analysis information, to regenerate analysis workflows, to detect errors and unusual behaviour in past analyses and to validate analyses. The provenance service has been based on the CRISTAL software [8], which is a data and workflow tracking (i.e. provenance) system. CRISTAL is a process modelling and provenance capture tool that addresses the harmonisation of processes by the use of a kernel so that multiple potentially heterogeneous processes can be integrated with each other and have their workflows tracked in the database. Using the facilities for description and dynamic modification in CRISTAL in a generic and reusable manner, CRISTAL is able to provide modifiable and reconfigurable workflows for a wide variety of Health (Grid) applications. CRISTAL also has pluggable data storage to store and retrieve provenance information and can be extended, to support a particular provenance database, by replacing its default storage mechanism.